\documentclass[twocolumn,aps,prl,reprint,amsmath,amssymb]{revtex4}

\pdfoutput=1
\usepackage{graphicx}
\usepackage{dcolumn}
\usepackage{bm}
\usepackage[utf8]{inputenc}
\usepackage{hyperref}

\begin{document}

\title{Extreme ultraviolet frequency comb metrology}

\author{Dominik Z. Kandula}
\author{Christoph Gohle}
\altaffiliation[Now at] { Ludwig-Maximilians-Universit\"at M\"unchen, Schellingstrasse 4, 80799 M
\"unchen}
\author{Tjeerd J. Pinkert}
\author{Wim Ubachs}
\author{Kjeld S.E. Eikema}
\email{KSE.Eikema@few.vu.nl}
\affiliation{Laser Centre Vrije Universiteit, De Boelelaan 1081, 1081HV Amsterdam, Netherlands}

\date{\today}

\begin{abstract}
The remarkable precision of frequency comb (FC) lasers is transferred to the extreme ultraviolet (XUV, 
wavelengths shorter than 100 nm), a frequency region previously not accessable to these devices. 
A frequency comb at XUV wavelengths near 51 nm is generated by amplification and coherent 
upconversion of a pair of pulses originating from a near-infrared femtosecond FC laser. The phase 
coherence of the source in the XUV is demonstrated using Helium atoms as a ruler and phase detector.
Signals in the form of stable Ramsey-like fringes with high contrast are observed when the FC laser is 
scanned over P states of Helium, from which the absolute transition frequency in the XUV can be 
extracted. This procedure yields a $^{4}$He ionization energy at $h\times5945204212(6)$~MHz, improved by
nearly an order of magnitude in accuracy, thus challenging QED calculations of this two-electron 
system.
\end{abstract}


\maketitle

Modelocked frequency-comb (FC) lasers~\cite{Holzwarth2000,Diddams20002} have revolutionized 
the field of precision laser spectroscopy. Optical atomic clocks using frequency combs are about to 
redefine the fundamental standard of frequency and time~\cite{Rosenband2008}. FC lasers have also 
vastly contributed to attosecond science by providing a way to synthesize electric fields at optical 
frequencies~\cite{Baltuska2003}, made long distance absolute length measurements possible~
\cite{Coddington2009}, and have recently been employed to produce ultracold molecules~
\cite{Ni2008}. FC based precision spectroscopy on simple atomic systems has provided one of the 
most stringent tests of bound state quantum electrodynamics (QED) as well as upper bounds on the 
drift of fundamental constants~\cite{Hansch2005}.  
Extending these methods into the extreme ultraviolet (XUV, wavelengths below 100 nm)  spectral 
region is highly desireable since this would for example allow novel precision QED tests~
\cite{Herrmann2009}.

Currently the wavelength range below 120~nm is essentially inaccessible to precision frequency 
metrology applications due to a lack of power of single frequency lasers and media for frequency 
upconversion. Spectroscopic studies on neutral Helium using amplified nanosecond laser pulses~
\cite{Eikema1997,Bergeson1998} are notoriously plagued by frequency chirping during amplification 
and harmonic conversion which limits the accuracy. 
These kind of transient effects can be avoided if a 
continuous train of high power laser pulses (produced by a FC) can be coherently upconverted. 
This would transfer the FC modes, at frequencies $f_{n} = f_{CEO} + n f_{rep}$, where $f_{CEO}$ 
is the carrier-envelope offset frequency, $f_{rep}$ is the repetition frequency of the pulses, and $n$ an 
integer mode number, to the XUV. Similar to what was shown in the 
visible~\cite{Marian2005,Fendel2007}, the upconverted pulse train could be used to directly 
excite a transition, with each of the upconverted modes acting like a single frequency laser.

By amplification of a few pulses from the train, and producing low harmonics in crystals and gasses, 
sufficient coherence has been demonstrated down to 125 nm to perform spectroscopic experiments~
\cite{Witte2005,Zinkstok2006a}. To reach wavelengths below 120 nm in the extreme ultraviolet or even 
x-rays, HHG has to be employed requiring nonlinear interaction at much higher intensities in the 
non-perturbative regime~\cite{Lewenstein1994}. That HHG can be phase coherent to some degree is 
known~\cite{Lewenstein1994,Zerne1997,Cavalieri2002}, and recently XUV light has been generated 
based on upconversion of all pulses of a comb laser at full repetition rate~
\cite{Jones2005,Gohle2005,Ozawa2008,Yost2009}. However, no comb structure in the harmonics has 
been demonstrated in the XUV, nor had these sources enough power to perform a spectroscopic 
experiment.

\begin{figure}
\centering
\includegraphics[width=8cm]{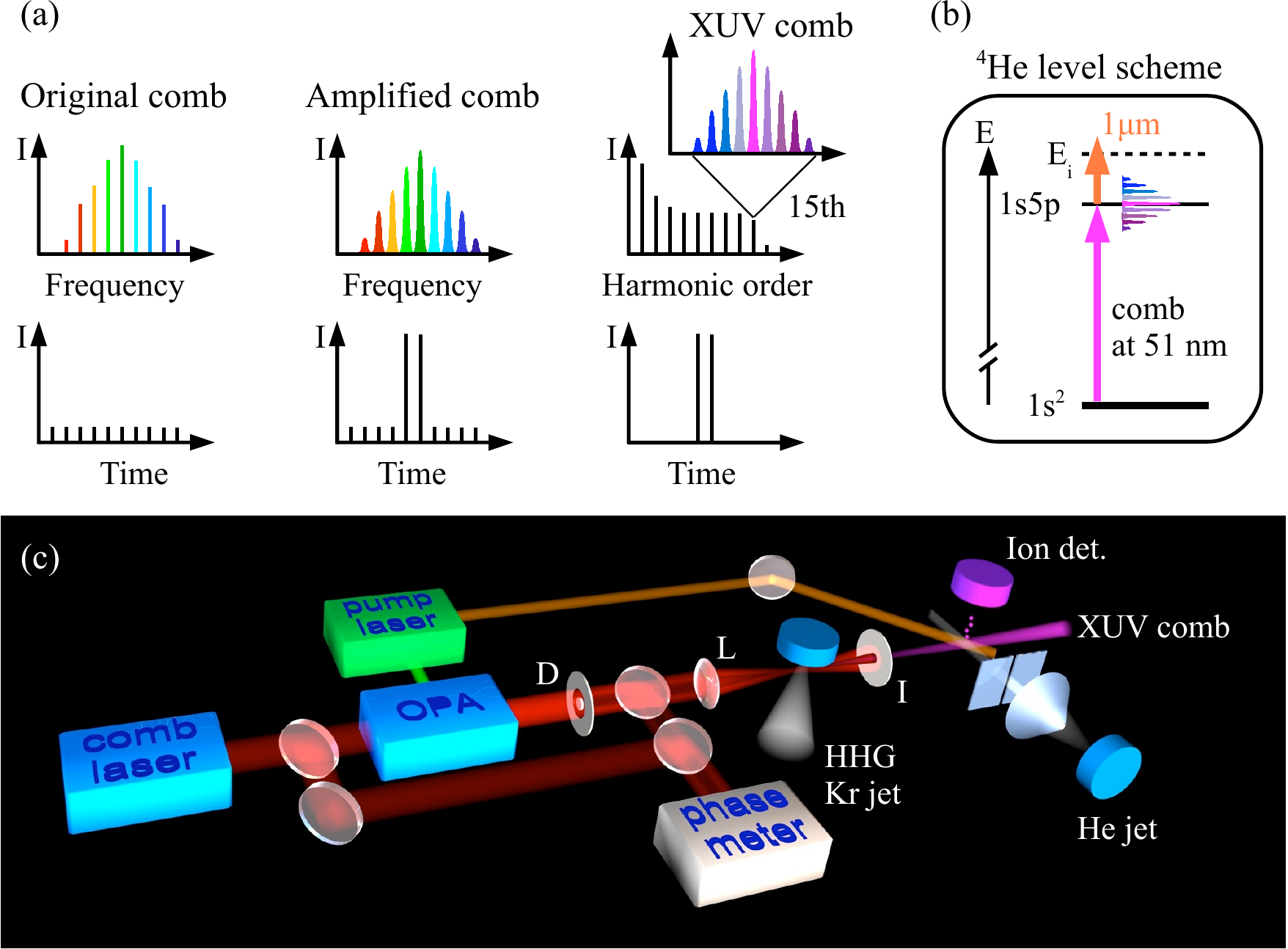}
\caption{
(a) Schematic of the 
spectral and temporal structure of the generated light at the different stages in the experiment (left to 
right): sharp equidistant frequencies from the infrared FC laser, cosine modulated spectrum after
double-pulse amplification, odd harmonics of the laser central frequency from HHG with cosine 
modulated XUV comb. (b) Simplified $^{4}$He level scheme,  XUV comb 
excitation at 51.5~nm from the $1s^{2}$ ground state to the $1s5p$ excited state and state selective
ionization by a pulse at 1064~nm. 
(c) Schematic of the 
experimental setup. D: beam mask, L: focusing lens, f=50 cm, I: iris to separate XUV from IR\label{fig:setup}}
\end{figure}

In this Letter we show that these limitations can be overcome, leading to the first absolute frequency 
measurement in the XUV.
Instead of converting a continuous train of FC pulses, we amplify a pair 
of subsequent pulses from a IR frequency comb laser with a double-pulse parametric amplifier 
(OPA)~\cite{Kandula2008} to the milli-joule level. These pulses with time separation $T=1/f_{rep}$ 
can be easily upconverted into the XUV with high efficiency using 
HHG in a dilute gaseous medium, and used to directly excite a transition in atoms or molecules (see 
fig. \ref{fig:setup}a and \ref{fig:setup}c). This form of excitation with two pulses resembles an optical 
(XUV) version of the Ramsey method of spatially (and therefore temporally) separated oscillatory 
fields~\cite{Ramsey1949,Witte2005}. In our case the fields are only separated in time. Excitation of an 
isolated (atomic or molecular) resonance with two (nearly) identical pulses produces a signal which is 
cosine-modulated according to $\cos(2\pi(f_{tr} T) - \Delta \phi(f_{tr}))$, where $f_{tr}$ is the transition 
frequency and $\Delta \phi(f_{tr})$ is the spectral phase difference at the transition frequency between 
the two pulses. Ideally, this spectral phase difference is just $\Delta \phi(f)= q\Delta\phi_{CE}=2q\pi 
f_{CEO}/f_{rep}$ where $q$ is the harmonic order under consideration and $\Delta\phi_{CE}$ the carrier 
envelope offset phase slip between subsequent pulses of the FC. In this case maxima of the cosine 
modulation lie at the locations where
one of the modes of an upconverted frequency comb would be resonant. This statement remains true even 
if the amplification and harmonic upconversion significantly distorts the electric field of the individual 
pulses as long as these distortions are common mode for each of them. Distortions that are not 
common mode need to be monitored and corrected for, in this experiment at a level of $< 1/200th$ of the 
fundamental IR field period. The frequency accuracy of the 
method is not fundamentally limited by the accuracy of this correction since an error $\delta$ in $\Delta 
\phi(f)$ translates into a frequency error $\Delta f = \delta/(2\pi T)$. Therefore the error can be made 
arbitrarily small by increasing the time separation $T$ between the pulses, provided the coherence 
time of the excited state allows this.

%

Phase coherent pulses near 773 nm are obtained from a Ti:Sapphire frequency comb (repetition rate 
$f_{rep}$ between 100 MHz and 185 MHz), which is linked to a GPS-controlled rubidium clock 
(Stanford research PRS10) to reach a stability on the order of $10^{-11}$ after a few seconds of 
averaging. A bandwidth of 6 nm (rectangular spectrum, leading to $\approx$300 fs pulses after 
compression) is selected from the FC laser so that after upconversion to  the XUV only one state in 
Helium is excited at a time. 
A non-collinear parametric double-pulse amplifier~\cite{Kandula2008} is used to amplify two 
subsequent FC pulses  (5.5--10 ns apart) at a repetition rate of 28 Hz. 
Parametric amplification intrinsically has small transient effects~\cite{Renault2007} so that differential 
pulse distortions are kept to a minimum. They are monitored using spectral interferometry with the unaltered original 
FC pulses as a reference~\cite{Kandula2008}. 
Wave front deformations in the beam are reduced by spatial filtering.  
Differential phase distortions from the amplifier and subsequent optics have a magnitude of typically 100~mrad in the IR. 
Spatial and spectral variations are at most 20-30~mrad. The IR beam is converted to a 
doughnut mode by a small disk mask (1.9 mm diameter compared to a beam diameter of 6 mm) to 
separate the XUV from the IR after HHG.
The remaining 1--2 mJ per pulse is focused in a pulsed krypton gas jet to $<5\times10^{13}$~ W$/$cm
$^2$ for HHG. An iris of 0.8 mm diameter placed at 40 mm distance after the focus allows the XUV to 
pass, but blocks the IR with a contrast ratio of 27:1.
For the 15th harmonic at 51.5 nm generated in the Krypton gas we estimate a yield of about $1 \times 
10^{8}$ photons per pulse. 

The resulting XUV beam intersects a low divergence beam of Helium atoms at perpendicular angle to 
avoid a Doppler shift. This beam is generated in a supersonic pulsed expansion (backing pressure 3 
bar) using a differential pumping stage containing two skimmers which limit the beam divergence to roughly 
3--4~mrad. This is similar to the divergence of the XUV beam ($<$2 mrad). The second skimmer 
position can be adjusted to set the XUV-He beam angle. To investigate Doppler effects, Helium can be seeded in heavier 
noble gases (partial pressure ratio 1:5). Pure Helium results in a velocity of 2000(315)~m/s, while seeding in Neon and Argon leads to a Helium velocity of 830(200) and 500(250)~m/s respectively.  

Helium atoms in the atomic beam are excited by the double pulse from the ground state into upper 
states which have spectral overlap with the HHG radiation (fig. \ref{fig:setup}b). 
After the double pulse has passed, the excited state population is determined by 
state-selective ionization of the Helium atoms using 60~ps, 1064~nm pulses from the OPA pump laser, 
followed by mass selective detection of the resulting ions in a time of flight spectrometer. Higher 
harmonics than the 15th are at least 10 times weaker, and produce a constant background signal of 
only 15\% of the relevant spectroscopic signal due to direct ionization. The 13th and lower order 
harmonics are not resonant with any transition from the ground state of Helium.



\begin{figure}
\centering
\includegraphics[width=8cm]{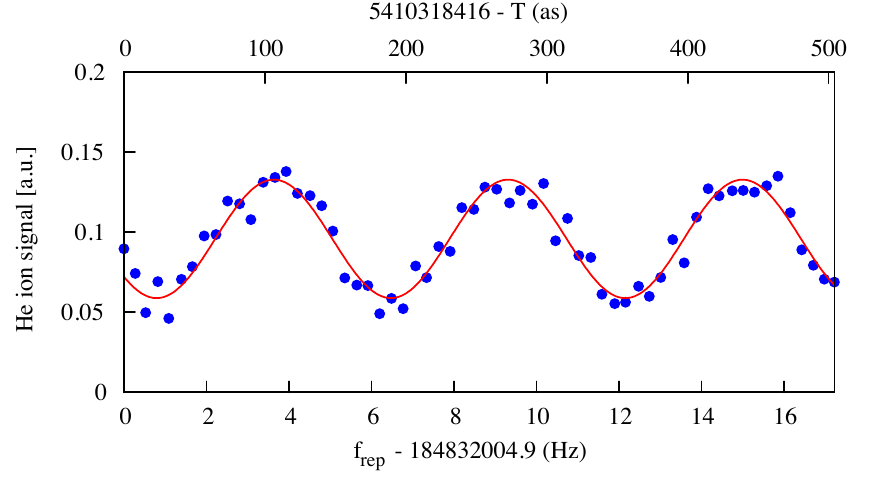}
\caption{Measured excitation probability (blue 
circles) of Helium at 51.5 nm on the 1s$^2$ $^1\mathrm{S}_0$ -- 1s5p$^1\mathrm{P}_1$ transition, normalized by the 
XUV pulse energy, as a function of the repetition rate $f_{rep}$ of the frequency comb laser. In this 
example $f_{CEO}$ is locked 46.21 MHz,
and a 1:5 He:Ne mixture is used for the atomic beam. The red line is a fit to the data.
\label{fig:he_signal_5p}}
\end{figure}

Figure~\ref{fig:he_signal_5p} shows a typical recording of the $^{4}$He ion signal, where the laser 
centre frequency is tuned to the 1s$^2$ $^1\mathrm{S}_0$ -- 1s5p $^1\mathrm{P}_1$
transition and the repetition frequency of the frequency comb is scanned. Recording such a trace takes 
about 20000 laser shots, corresponding to ten minutes of continuous data taking. During a recording 
the pulse delay $T$ is changed in steps of typically less than 1 attosecond every 28 laser shots, with a 
total change over a scan of less than 1 femtosecond. 

By fitting the phase of the expected cosine function to this signal, we determine the transition 
frequency up to an integer multiple of the laser repetition 
frequency $f_{rep}$. The statistical error in the fit of a single recording is typically $1/50$th of a 
modulation period.
Depending on the repetition rate it amounts to a uncertainty of 2-3~MHz in the observed transition 
frequency, which is unprecedented in the XUV spectral region. Such scans are repeated many times
to assess systematic effects.
The dominant systematic shifts are Doppler shift and a differential phase shift of the XUV pulses 
due to changing ionization of the HHG medium. The former is minimized by setting the XUV-Helium beam angle
perpendicular. It is evaluated by varying the speed of the Helium beam and extrapolating the transition 
frequency to zero velocity. The ionization shift is found by varying the HHG medium density (assuming
ionized fraction remains constant) and extrapolating to zero density. The statistical error in these 
extrapolations dominates the final statistical error. Systematic errors in Doppler shift and ionization shift
originate in the uncertainty of the Helium velocity and ionization dynamics in the HHG medium, 
respectively. 
Other effects that are taken into account include recoil shifts, refractive index changes (Kerr effect) in 
the focusing lens for HHG and the entrance  window to the vacuum setup, AC and DC Stark effect and 
Zeeman shift. A summary of the error budget can be found in table \ref{table:errors}.
Most recordings were made on the 1s$^{2}~^{1}\mathrm{S}_{0}$--1s5p $^{1}\mathrm{P}_{1}$ transition at 51.5~nm. As a 
cross check also a series was measured on the 1s$^{2}~^{1}\mathrm{S}_{0}$--1s4p $^{1}\mathrm{P}_{1}$ transition at 
52.2~nm.
The $^{4}$He ionization potential (up to an integer multiple of $f_{rep}$) is derived from these 
measurements by adding the excited state ionization energy of the 4p and 5p.
The energy of these states is known with an accuracy better 
than 20 kHz based on theoretical calculations~\cite{Morton2006}. 

\begin{figure}
\centering
\includegraphics[width=8cm]{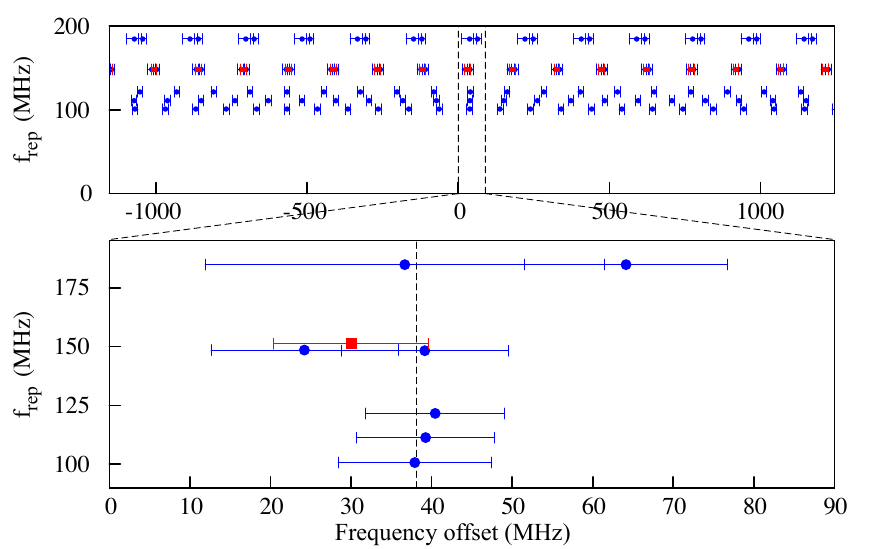}
\caption{(top) $^{4}$He ground state ionization energy $\pm n\times f_{rep}$ based on the 
$1s^{2}\ ^{1}\mathrm{S}_{0}-1s5p ^{1}\mathrm{P}_{1}$ (blue circles) and the 
$1s^{2}\ ^{1}\mathrm{S}_{0}-1s4p ^{1}\mathrm{P}_{1}$ transition (red squares). (bottom) Zoom at the 
coincidence point for all repetition frequencies. The vertical line at +38(6)~MHz represents the
weighted mean. 
All values are relative to the theoretical value of 5945204174 MHz~\cite{Yerokhin2010}. (square point slightly shifted up for visibility) \label{fig:vernier}}
\end{figure}

\begin{table}
\begin{tabular}{lr}
statistical error	& 3.7~MHz \\ 
Ionization shift	& $4.9$~MHz  \\
Doppler shift	& $500$~kHz \\
DC-Stark shift	& $< 1$~kHz  \\
signal from other levels & $< 30$~kHz \\
Zeeman shift	& $<7$~kHz  \\ 
\hline
Total			& $6$~MHz
\end{tabular}
\caption{The major contributions to the error budget of the ionization potential.}\label{table:errors}
\end{table}
To remove the ambiguity due to the periodic comb spectrum, we repeated this procedure for several 
repetition frequencies within the range of 100~MHz and 185~MHz.
The correct ``mode number'' is found by plotting the possible ionization energies of the Helium ground 
state against $f_{rep}$ as shown in figure \ref{fig:vernier}. 
A clear coincidence between the results for different repetition rates can be seen, leading to a new 
ground ionization energy for $^{4}$He of 5945204212(6) MHz by taking a weighted average over all 
measured frequencies at the coincidence location. 
This is in agreement with recent theoretical predictions of 5945204174(36)~MHz~\cite{Yerokhin2010} 
and 5945204175(36)~MHz~\cite{Drake2008} within the combined uncertainty of theory and 
experiment. Compared to previous experiments employing single 
nanosecond duration laser pulses, we find good agreement with the value of 5945204215(45)~MHz~
\cite{Eikema1997} (using the most recent 2p state ionization energy~\cite{Yerokhin2010}, and 
corrected for a 14.6 MHz recoil shift that was previously not taken into account). However, there is a 
difference of nearly 3$\sigma$ compared to a competing result~\cite{Bergeson1998}. 

From the observed signal contrast (defined as the modulation amplitude divided by the average signal 
level) we can infer the temporal coherence of the HHG process relevant for spectroscopy experiments 
as well as attosecond physics. The contrast depends on several parameters: the upper state lifetime 
(natural linewidth), the time between the pulses, a difference in XUV pulse energy of the two pulses, 
the frequency stability of the interference pattern, Doppler broadening of the transition, and the 
previously mentioned constant background from direct ionization due to the 17th and higher 
harmonics. The Doppler broadening is dominated by the effective atomic beam opening angle and the 
radial velocity of the beam. All these effects lead to a varying contrast depending on the Helium velocity $v$ and comb 
repetition frequency. For a high repetition rate ($f_{rep}=185$~MHz) and low $v$ (Helium seeded in argon), 
we find a fringe contrast of 55\%, while on the other hand for $f_{rep}=100$~MHz 
and a pure Helium beam (large $v$) the contrast is below 5\%. From these observed 
variations and a straight forward model for the visibility as a function of $f_{rep}$ and atomic beam velocity, 
we estimate 
a phase jitter of 0.38(6) cycles in the XUV. To a large part (0.21 cycles), this instability can be attributed to noise of 
the driving IR pulses. This noise in turn is 
dominated by frequency noise of the frequency comb laser itself. It can be reduced 
significantly by locking the FC to a low noise optical reference resonator, instead of the current $f_{rep}
$ lock which is based on a radio frequency reference only. 

In conclusion, we have demonstrated frequency comb generation in the XUV and performed the first 
absolute frequency determination in this spectral region. Based on the contrast of the Helium excitation 
signal we find that the excess  phase noise in the HHG process used to generate the XUV comb is at 
most 0.3 optical cycles in the XUV. This means that the timing of the generated electric field of the XUV 
waveform of individual pulses is stable within less than 50~as, which is an important benchmark for 
both spectroscopy applications as well as ultrafast physics. 
The new value of the $^{4}$He ionization potential is in good agreement with theory~
\cite{Yerokhin2010,Drake2008} and already almost an order of magnitude more accurate than the best 
previous results using single nanosecond laser pulses~
\cite{Eikema1997,Bergeson1998}.
Moreover, the accuracy of our method can readily be improved by orders of magnitude by increasing 
the time delay between the two pulses. One could, for example, perform high resolution spectroscopy 
on the 1s-2s two photon transition of hydrogen-like Helium ions at 60~nm, which is very promising to 
perform QED tests beyond what has been possible so far in atomic hydrogen~
\cite{Fischer2004,Herrmann2009}.
The results show that as long as the carrier phase noise is kept low enough not to destroy the mode 
structure, comb generation should be extendable to the soft X-ray region. This may allow applications 
such as coherent XUV/X-ray imaging, precision QED tests of (highly) charged ions, to perhaps 
ultimately x-ray nuclear clocks.

This work was supported by the Foundation for Fundamental Research on Matter (FOM) through its IPP-program `Metrology with frequency comb lasers', by the Netherlands Organisation for Scientific Research (NWO), by the EU via LASERLAB-Europe and via the Early Stage Training network 'Atlas', and by the 
Humboldt Foundation. We thank W.~Vassen, R.~van Rooij and J.~Simonet for sharing their Er:fiber laser with us.

\bibliographystyle{apsrev}
\bibliography{paper}

\end{document}